\documentclass[journal]{IEEEtran}
\IEEEoverridecommandlockouts
\usepackage{cite}
\usepackage{amsmath,amssymb,amsfonts}
\usepackage{algorithmic}
\usepackage{graphicx}
\usepackage{textcomp}
\usepackage{xcolor}
\usepackage{array}

\def\BibTeX{{\rm B\kern-.05em{\sc i\kern-.025em b}\kern-.08em
    T\kern-.1667em\lower.7ex\hbox{E}\kern-.125emX}}
\begin{document}

\title{Survey and Benchmarking of Machine Learning Accelerators\\
\thanks{This material is based upon work supported by the Assistant Secretary of Defense for Research and Engineering under Air Force Contract No. FA8721-05-C-0002 and/or FA8702-15-D-0001. Any opinions, findings, conclusions or recommendations expressed in this material are those of the author(s) and do not necessarily reflect the views of the Assistant Secretary of Defense for Research and Engineering.}
}

\author{\IEEEauthorblockN{Albert Reuther, Peter Michaleas, Michael Jones, Vijay Gadepally, Siddharth Samsi, and Jeremy Kepner} \\
\IEEEauthorblockA{\textit{MIT Lincoln Laboratory Supercomputing Center} \\
Lexington, MA, USA \\
\{reuther,pmichaleas,michael.jones,vijayg,sid,kepner\}@ll.mit.edu}
}

\maketitle

\begin{abstract}

Advances in multicore processors and accelerators have opened the flood gates to greater exploration and application of machine learning techniques to a variety of applications. These advances, along with breakdowns of several trends including Moore's Law, have prompted an explosion of processors and accelerators that promise even greater computational and machine learning capabilities. These processors and accelerators are coming in many forms, from CPUs and GPUs to ASICs, FPGAs, and dataflow accelerators. 

This paper surveys the current state of these processors and accelerators that have been publicly announced with performance and power consumption numbers. The performance and power values are plotted on a scatter graph and a number of dimensions and observations from the trends on this plot are discussed and analyzed. For instance, there are interesting trends in the plot regarding power consumption, numerical precision, and inference versus training. We then select and benchmark two commercially-available low size, weight, and power (SWaP) accelerators as these processors are the most interesting for embedded and mobile machine learning inference applications that are most applicable to the DoD and other SWaP constrained users. We determine how they actually perform with real-world images and neural network models, compare those results to the reported performance and power consumption values and evaluate them against an Intel CPU that is used in some embedded applications. 

\end{abstract}

\begin{IEEEkeywords}
Machine learning, GPU, TPU, dataflow, accelerator, embedded inference
\end{IEEEkeywords}

\section{Introduction}

\begin{figure}[th]
    \centering
    \includegraphics[width=3in]{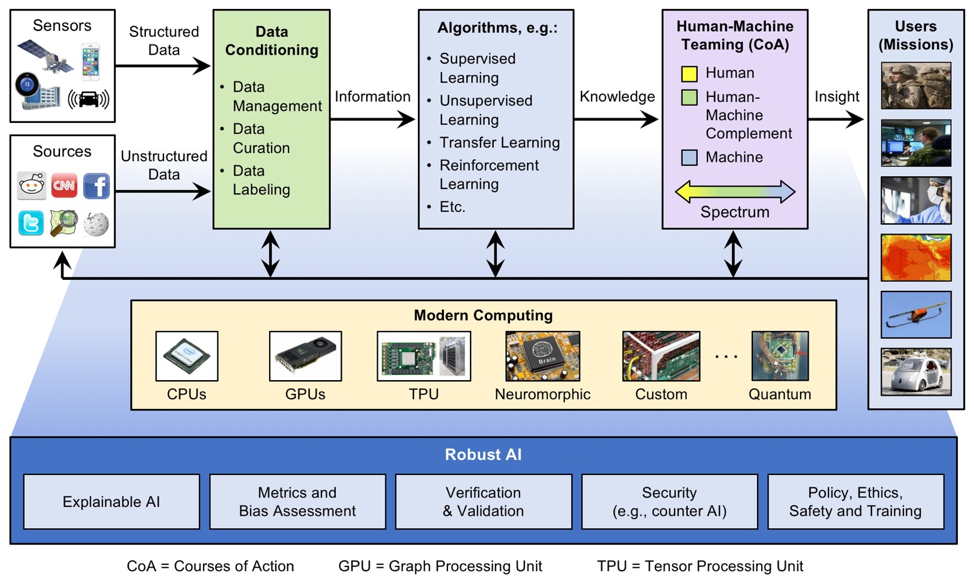}
    \caption{Canonical AI architecture consists of sensors, data conditioning, algorithms, modern computing, robust AI, human-machine teaming, and users (missions). Each step is critical in developing end-to-end AI applications and systems.}
    \label{fig:architecture}
  \end{figure}

Artificial Intelligence (AI) and machine learning (ML) have the opportunity to revolutionize the way many industries, militaries, and other organizations address the challenges of evolving events, data deluge, and rapid courses of action. Innovations in computations, data sets, and algorithms have driven many advances for machine learning and its application to many different areas. AI solutions involve a number of different pieces that must work together in order to provide capabilities that can be used by decision makers, warfighters, and analysts; Figure~\ref{fig:architecture} depicts these important pieces that are needed when developing an end-to-end AI solution. While certain components may not be as visible to end-users as others, our experience has shown that each of these interrelated components play a major role in the success or failure of an AI system. 

On the left side of Figure~\ref{fig:architecture}, structured and unstructured data sources provide different views of entities and/or phenomenology. These raw data products are fed into a data conditioning step in which they are fused, aggregated, structured, accumulated, and converted to information. The information generated by the data conditioning step feeds into a host of supervised and unsupervised algorithms such as neural networks, which extract patterns, predict new events, fill in missing data, or look for similarities across datasets, thereby converting the input information to actionable knowledge. This actionable knowledge is then passed to human beings for decision-making processes in the human-machine teaming phase. The phase of human-machine teaming provides the users with useful and relevant insight turning knowledge into actionable intelligence or insight. 

Underlying all of these phases is a bedrock of modern computing systems that is comprised of one or more heterogenous computing elements. For example, sensor processing may occur on low power embedded computers, while algorithms may be computed in very large data centers. With regard to performance advances in these computing elements, Moore's law trends have ended~\cite{theis2017end}, as have a number of related laws and trends including Denard's scaling (power density), clock frequency, core counts, instructions per clock cycle, and instructions per Joule (Koomey's law)~\cite{horowitz2014computing}. Many of the technologies, tricks and techniques of processor chip designers that extended these trends have been exhausted. However, all is not lost, yet; advancements and innovations are still progressing. In fact, there has been a Cambrian explosion of computing technologies and architectures in recent years. Specialization of circuits for certain functionalities is being exploited whereby certain often-used operational kernels, methods, or functions are being accelerated with specialized circuit blocks and chips.  These accelerators are designed with a different balance between performance and functional flexibility. One area in which we are seeing an explosion of accelerators is ML processors and accelerators~\cite{hennessy2019new}. Understanding the relative benefits of these technologies is of particular importance to applying AI to domains under significant constraints such as size, weight, and power, both in embedded applications and in data centers. 

But before we get to the survey of ML processors and accelerators, we must cover several topic that are important for understanding several dimensions of evaluation in the survey. We must discuss the types of neural networks for which these ML accelerators are being designed; the distinction between neural network training and inference; and the numerical precision with which the neural networks are being used for training and inference: 

\begin{itemize}

\item Types of Neural Networks -- AI and machine learning encompass a wide set of statistics-based technologies as one can see in the taxonomy detailed in the algorithm section (Section 3) of this MIT Lincloln Laboratory technical report~\cite{gadepally2019enabling}. Even among neural networks, there are a growing number of neural network patterns~\cite{vanveen2019neural}. This paper will focus on processors that are geared toward deep neural networks (DNNs) and convolutional neural networks (CNNs).  Overall, the most emphasis of computational capability for machine learning is on DNN and CNNs because they are quite computationally intensive~\cite{canziani2016analysis}, with the fully connected and convolutional layers being the most computationally intense. Conversely, pooling, dropout, softmax, and recurrent/skip connection layers are not computationally intensive since these types of layers stipulate datapaths for weight and data operands. 

\item Neural Network Training versus Inference -- Neural network training uses libraries of input data to converge model weight parameters by applying the labeled input data (forward projection), measuring the output predictions and then adjusting the model weight parameters to better predict output predictions (back projections).  Neural network inference is using a trained model of weight parameters and applying it to input data to receive output predictions. Processors designed for training can also perform well at inference, but the converse is not always true. 

\item Numerical precision -- The numerical precision with which the model weight parameters are stored and computed has an impact on the effectiveness and efficiency with which networks are trained and used for inference. Generally higher numerical precision representations, particularly floating point representations, are used for training, while lower numerical precision representations, including integer representations, have been shown to be reasonably effective for inference~\cite{sze2017efficient, narang2018mixed}. However, it has also generally been established that very limited numerical precisions like int4, int2, and int1 do not adequately represent model weight parameters and significantly affect model output predictions. 


\end{itemize}

The survey in the next section of this paper focuses on the computational throughput of the processors and accelerators along with the power that is consumed to achieve that performance. Other factors include the memory bandwidth to load and update model parameters and data; memory capacity for model weight parameters and input data, both close to the arithmetic units and the global memory of the processors and accelerator; and arithmetic intensity~\cite{williams2009roofline} of the neural network models being processed by the processor or accelerator. These factors are involved in managing model parameters and input data flows within the model; hence, they also influence the trade-offs between chip bandwidth capabilities, data flow flexibility, and configuration and amount of computational capability. These factors, however, are beyond the scope of this paper, and they will be addressed in future phases of this research. 

\section{Survey of Processors}

Many recent advances in AI can be at least partly credited to advances in computing hardware~\cite{krizhevsky2012imagenet,jouppi2018domain}. In particular, modern computing advances have been able to realize many computationally heavy machine-learning algorithms such as neural networks. While machine-learning algorithms such as neural networks have had a rich theoretic history~\cite{minsky1967computation}, recent advances in computing have made the application of such algorithms a reality by providing the computational power needed to train and process massive quantities of data. Although the computing landscape of the past decade has been rich with numerous innovations, more embedded and mobile applications that require low size, weight, and power (SWaP) systems will need capabilities that are beyond those delivered by the traditional architectures of central processing units (CPUs) and graphics processing units (GPUs). For example, in commercial applications, it is common to off-load data conditioning and algorithms to non-SWaP constrained platforms such high-performance computing clusters or processing clouds. Defense applications among others, on the other hand, may need AI applications to be performed inside low-SWaP platforms or local networks (edge computing) and without the use of the cloud due to insufficient security or communication infrastructure. 

\begin{figure*}[htb]
    \includegraphics[width=\textwidth]{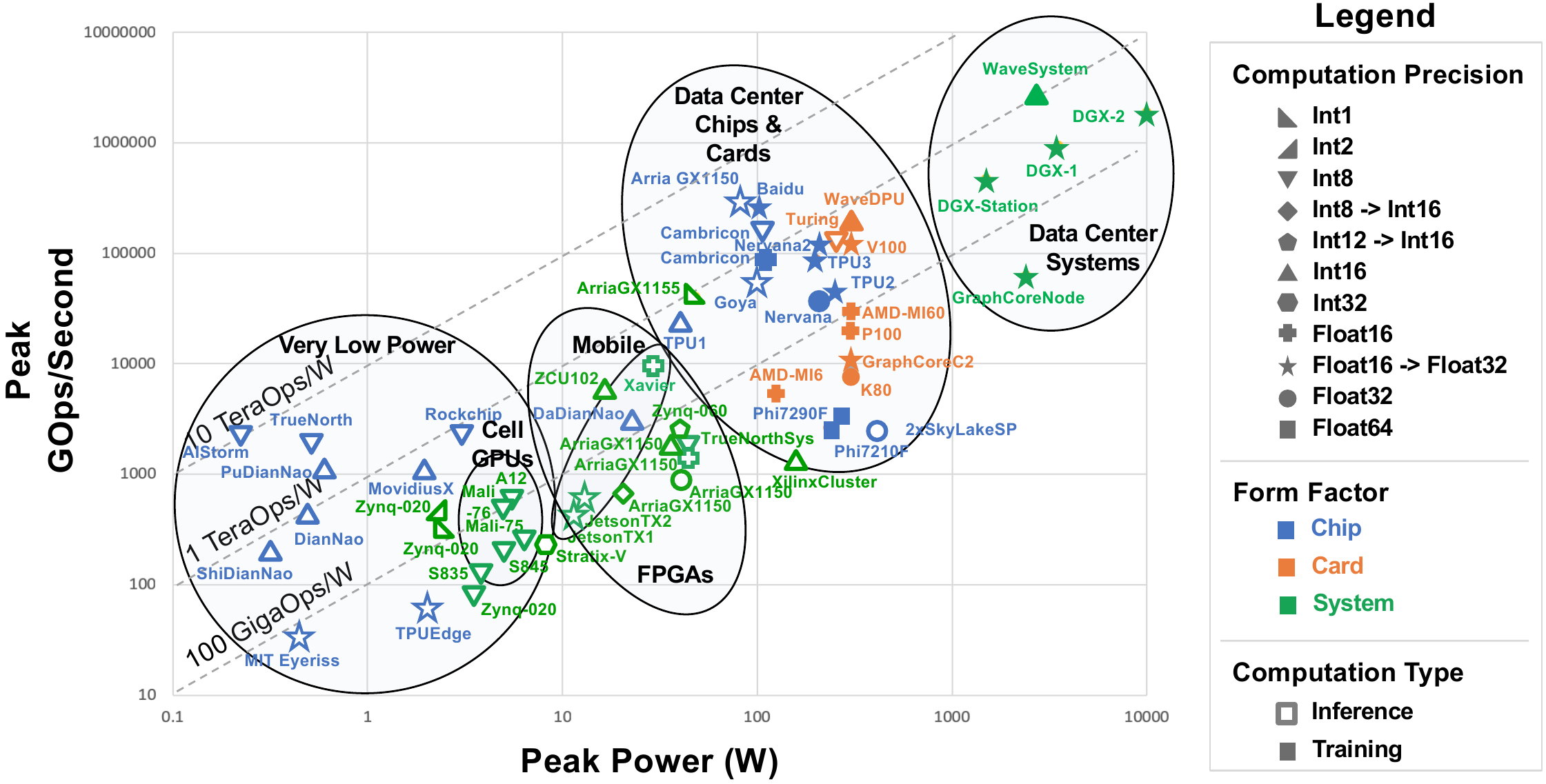}
    \caption{Performance vs. power scatter plot of publicly announced AI accelerators and processors.}
    \label{fig:PerformancePower}
  \end{figure*}

The survey in this section gathers performance and power information from publicly available materials including research papers, technical trade press, company benchmarks, etc. While there are ways to access information from companies and startups (including those in their silent period), this information is intentionally left out of this survey; such data will be included in this survey when it becomes publicly available. The key metrics of this public data are plotted in Figure~\ref{fig:PerformancePower}, which graphs recent processor capabilities (as of May 2019) mapping peak performance vs. power consumption. 
The x-axis indicates peak power, and the y-axis indicate peak giga operations per second. (GOps/s) Note the legend on the right, which indicates various parameters used to differentiate computing techniques and technologies. The computational precision of the processing capability is depicted by the geometric shape used; the computational precision spans from single bit int1 to single byte int8 and four-byte float32 to eight-byte float64. The form factor is depicted by the color; this is important for showing how much power is consumed, but also how much computation can be packed onto a single chip, a single PCI card, and a full system. Blue is only the performance and power consumption of a single chip. Orange shows the performance and power of a card (note that they all are in the 200-300 Watt zone). Green shows the performance and power of entire systems -- in this case, single node desktop and server systems. This survey is limited to single motherboard, single memory-space systems. Finally, the hollow geometric objects are performance for inference only, while the solid geometric figures are performance for training (and inference) processing. Mostly, low power solutions are only capable of inference, though there are some high-power accelerators (WaveDPU, Goya, Arria, and Turing) that are targeting high performance for inference only.

From Figure~\ref{fig:PerformancePower}, we can make a number of general observations. First, much of the recent efforts have focused on processors that are in the 10-300W range in terms of power utilization, since they are being designed and deployed as processing accelerators. (300W is the upper limit for a PCI-based accelerator card.) For this power envelope, the performance can vary depending on a variety of factors such as architecture, precision, and workload (training vs. inference). There are many solutions under the 1 TeraOps/W line; however, there are several inference solutions and a few training solutions that are reporting greater than 1 TeraOps/W. 

With the current offerings, at least 100W must be employed to perform training; all of the points on the scatter plot below 100W are inference-only processors/accelerators. There are a number of possible explanations for this, but it is likely that there is currently little driving a requirement for low-power training, though there is much demand for low-power inference on devices ranging from smartphones to remotely piloted aircraft (RPA) and autonomous vehicles. From a technology standpoint, this may suggests that the trade-offs necessary to do neural network training under the 100W envelope affect the performance, numerical accuracy, and prediction accuracy too greatly. 

Many hardware manufacturers, faced with limitations in fabrication processes, have been able to exploit the fact that machine-learning algorithms such as neural networks can perform well even when using limited or mixed precision~\cite{narang2018mixed,gupta2015deep} representation of activation functions, weights, and biases. Such hardware platforms (often designed specifically for inference) may quantize weights and biases to half precision (16 bits) or even single bit representations in order to improve the number of operations/second without significant impact to model prediction, accuracy, or power utilization. To that point, in the inference engines, the entire neural network model is usually loaded onto the chip before any inference is performed. Loading the model turns the model's parameters into constants that are stored with the operator rather than operands that must be loaded from volatile (DRAM or SRAM) memory thereby reducing the number of operand/parameter loads that must occur separate from the instruction load.

There are a number of dimensions with which we can present the processors and accelerators in this survey. We have chosen to roughly categorize the scatter plot into six regions that roughly correspond to performance and power consumption: Very Low Power and Research Chips, Cell (Smartphone) GPUs, Mobile and Embedded Chips and Systems, FPGA Accelerators, Data Center Chips and Cards, and Data Center Systems. In the following listings, the angle-bracketed string is the label of the item on the scatter plot, and the square bracket after the angle bracket is literature reference from which the performance and power values came. Some of the performance values are reported in frames per second (fps) with a given machine learning model. For those values, Samuel Albanie has Matlab code and a web site that lists all of the major machine learning models with their operations per epoch/inference, parameter memory, feature memory, and input size~\cite{albanie2019convnet}; the operations per epoch/inference are used to compute operations per second from frames per second. Finally, if a neural network model is not mentioned, the performance reported is peak performance. 

\subsection{Very Low Power and Research Chips}

Chips in the very low power regime have been mainly university and industry research chips. However, a few vendors have announced or are offering products in this space.  

\begin{itemize}
\item MIT Eyeriss chip $\langle$Eyeriss$\rangle$~\cite{chen2018eyeriss,chen2017eyeriss,sze2017efficient} is a research chip from Vivienne Sze's group in MIT CSAIL. Their goal was to develop the most energy efficient inference chip possible. The result was acquired running AlexNet with no mention of batch size. 
\item The TrueNorth $\langle$TrueNorth$\rangle$~\cite{akopyan2015truenorth,feldman2016ibm} is a digital neuromorphic research chip from the IBM Almaden research lab. It was developed under DARPA funding in the Synapse program to demonstrate the efficacy of digital spiking neural network (neuromorphic) chips. Note that there are points on the graph for both the system, which draws the 44 W power, and the chip, which itself only draws up to 275 mW. 
\item The Intel MovidiusX processor $\langle$MovidiusX$\rangle$~\cite{hruska2017new} is an embedded video processor that includes a Neural Engine for video processing and object detection. 
\item In early 2019, Google released a TPU Edge processor $\langle$TPUEdge$\rangle$~\cite{tpu2019edge} for embedded inference application. The TPU Edge uses TensorFlow Lite, which encodes the neural network model with low precision parameters for inference.
\item The DianNao series of dataflow research chips came from a university research team in China. They published four different designs aimed at different types of ML processing~\cite{chen2016diannao}. The DianNao $\langle$DianNao$\rangle$~\cite{chen2016diannao} is a neural network inference accelerator, and the DaDianNao $\langle$DaDianNao$\rangle$~\cite{chen2014dadiannao} is a many-tile version of the DianNao for larger NN model inference. The ShiDianNao $\langle$ShiDianNao$\rangle$~\cite{du2015shidiannao} is designed specifically for convolutional neural network inference. Finally, the PuDianNao $\langle$PuDianNao$\rangle$~\cite{liu2015pudiannao} is designed for seven representative machine learning techniques: k-means, k-NN, na\"ive Bayes, support vector machines, linear regression, classification tree, and deep neural networks. 
\item San Jose startup AIStorm $\langle$AIStorm$\rangle$~\cite{merrit2019startup} claims to do some of the math of inference up at the sensor in the analog domain. They originally came to the embedded space scene with biometric sensors and processing. They call their chip an AI-on-Sensor capability. 
\item The Rockchip RK3399Pro $\langle$Rockchip$\rangle$~\cite{rockchip2018rockchip} is an image and neural co-processor from Chinese company Rockchip. They published raw performance numbers for 8bit inference. This appears to be a GPU-based co-processor but details are few. 
\end{itemize}

\subsection{Cell / Smartphone GPU-based Neural Engines}
A number of smartphone vendors are embedding GPU-based neural engines in their smartphones to enable object detection, face recognition, and other inference-based tasks. The  performance metrics for five inference neural engines, which were benchmarked with AImark, are included in this survey. AImark runs VGG-16, ResNet34 and InceptionV3 on smartphones, and it is available in the Apple App Store and the Google Play Store. It is reasonably safe to assume that these GPU-based vector processors are executing with Int8 precision. 
\begin{itemize}
\item The Apple A12 processor $\langle$A12$\rangle$~\cite{frumusanu2018iphone,peng2018ai} in the iPhone Xs tops out this set. This A12 neural engine bursts its power utilization to 5.5W for short time periods (above its usually 5W maximum for battery life) for fast inference runs, and this performance point is on the VGG-16 model. 
\item The Huawei Kirin 980 (with AMD Mali-76 GPU IP) $\langle$Mali-76$\rangle$~\cite{frumusanu2018samsung} and Kirin 970 (with AMD Mali-75 GPU IP) $\langle$Mali-75$\rangle$~\cite{frumusanu2018hisilicon} make their performance mark with the ResNet34 and VGG-16 models, respectively. 
\item Finally, the Qualcomm Snapdragon 835 $\langle$S835$\rangle$ and 845 $\langle$S845$\rangle$~\cite{frumusanu2018samsung} are also on the chart with performance numbers using the ResNet34 and InceptionV3 models, respectively.
\end{itemize}

\subsection{Embedded Chips and Systems} 

The systems in this category are aimed at automotive AI/ML, autonomous vehicles, UAVs, robots, etc. They all have several ARM cores that are mated with NVIDIA CUDA GPU cores. 

\begin{itemize}
\item The NVIDIA Jetson-TX1 $\langle$JetsonTX1$\rangle$~\cite{franklin2017nvidia} incorporates 4 ARM cores and 256 CUDA Maxwell cores. It is aimed at low power applications for inference only. The performance was achieved with GoogLeNet with a batch size of 128. 
\item The Jetson-TX2 $\langle$JetsonTX2$\rangle$~\cite{franklin2017nvidia} mates 6 ARM cores with 256 CUDA Pascal cores. It also is aimed at low power applications for inference only. The performance was achieved with GoogLeNet with a batch size of 128. 
\item The NVIDIA Xavier $\langle$Xavier$\rangle$~\cite{hruska2018nvidia} deploys 8 ARM cores with 512 CUDA Volta cores and 64 Tensor cores. It is aimed also at low power applications for inference only. 
\end{itemize}

\subsection{FPGA Co-processors}

In public literature, the use of FPGAs for neural networks has been primarily in the technical research domain. Quite a number of research teams around the world have mapped one or more neural network models onto one or more FPGAs and collected a variety of performance and model prediction accuracy metrics. Several survey papers have been published including~\cite{li2017survey} and~\cite{mittal2018survey}, and the most comprehensive survey paper of mapping and running DNNs on FPGAs is here~\cite{guo2017survey}. This last paper lists 25 top results from published research literature, of which we have chosen 12 that are the performance leaders for their numerical precision and/or FPGA model. They are labeled with an abbreviation of their chip type: $\langle$Zynq-020$\rangle$ int1~\cite{nakahara2017fully}, int2~\cite{jiao2017accelerating}, int8~\cite{guo2018angeleye}; $\langle$Zynq-060$\rangle$ int16 accumulator/int12 result~\cite{han2017ese}, $\langle$ZCU102$\rangle$ int16~\cite{lu2017evaluating}, $\langle$Stratix-V$\rangle$ int32~\cite{podili2017fast}, $\langle$ArriaGX1150$\rangle$ int16 accumulator/int8 result~\cite{ma2017optimizing}, int16~\cite{zhang2017improving}, fp16~\cite{aydonat2017opencl}, fp32~\cite{zhang2017improving}; and $\langle$ArriaGX1155$\rangle$ 1-bit~\cite{moss2017high} points with different numerical precisions. They are all used for inference. Finally, there is a 7-FPGA Xilinx Cluster $\langle$XilinxCluster$\rangle$~\cite{zhang2016energy} in which the research team ganged together one control FPGA and six computational FPGAs to execute much larger neural network models. All of these results are from running one of the following models: AlexNet, VGG-16, VGG-19, DoReFa-Net, and an LSTM model. Details are in~\cite{guo2017survey}. 

\subsection{Data Center Chips and Cards}

There are a variety of technologies in this category including several CPUs, a number of GPUs, a CPU-controlled FPGA solution, and dataflow accelerators. They are addressed in their own subsections to group similar processing technologies. 

\subsubsection{CPU-based Processors}
\begin{itemize}
\item The Intel SkyLake SP processors $\langle$2xSkyLakeSP$\rangle$~\cite{rodriguez2017intel,intel2019platnum} are conventional Xeon server processors. Intel has been marketing these chips to data analytics companies as very versatile inference engines with reasonable power budgets. The performance numbers were measured using Caffe ResNet-50 with batch size of 64 on a 2-socket SkyLakeSP system. 
\item The Intel Xeon Phi processor chips have 64, 68, or 72 cores, with each core having four hardware hyper-threads and two AVX-512 (512-bit wide) vector units~\cite{jeffers2016intel}. Having these 128 AVX-512 vector units on a 64-core chip is equal to 2048 double precision floating point vector ALUs or 4096 single precision floating point vector ALUs. The Phi7210F $\langle$Phi7210F$\rangle$~\cite{wikipedia2019xeonphi} is the 64-core chip we have in the TX-Green Petaflop system, while the Phi7290F $\langle$Phi7290F$\rangle$~\cite{wikipedia2019xeonphi} is the top bin, 72-core Xeon Phi (KNL). 
\end{itemize}

\subsubsection{CPU-Controlled FPGA}

The Intel Arria solution pairs an Intel Xeon CPU with an Altera Arria FPGA $\langle$Arria GX1150$\rangle$~\cite{hemsoth2018intel,abdelfattah2018dla} (next to the Baidu point). The CPU is used to rapidly download FPGA hardware configurations to the Arria, and then farms out the operations to the Arria for processing certain key kernels. Since inference models do not change, this technique is well geared toward this CPU-FPGA processing paradigm. However, it would be more challenging to farm ML model training out to the FPGAs. The performance benchmark was on an Arria 10 1150 FPGA using GoogLeNet reporting 900 fps.

\subsubsection{GPU-based Accelerators}

There are four NVIDIA cards and two AMD/ATI cards on the chart (listed respectively): the Maxwell architecture K80 $\langle$K80$\rangle$~\cite{smith2014nvidia}, the Pascal architecture P100$\langle$P100$\rangle$~\cite{pascal2018nvidia,smith2016nvidia}, the Volta architecture V100 $\langle$V100$\rangle$~\cite{volta2019nvidia,smith201816gb}, the TU106 Turing $\langle$Turing$\rangle$~\cite{kilgariff2018nvidia}, the MI6 $\langle$MI6$\rangle$~\cite{exxactcorp2017taking}, and MI60 $\langle$MI60$\rangle$~\cite{smith2018amd}. The K80, P100, V100, MI6, and MI60 GPUs are pure computation cards intended for both inference and training, while the TU106 Turing GPU is geared to the gaming/graphics market for including inference processing within the graphics processing.

\subsubsection{Data Center Chips and Cards}

This subsection lists a series of chips and cards intended for data center deployment. 

\begin{itemize}

\item Intel Corp. bought AI chip startup Nervana in August 2016 to enter the AI accelerator market. The first Nervana chip $\langle$Nervana$\rangle$~\cite{rao2018beyond} called Lake Crest is scheduled to ship in 2019. The follow-on is called Spring Crest $\langle$Nervana2$\rangle$~\cite{rao2018beyond}, and it is scheduled to ship in late 2019.

\item Google has released three versions of their Tensor Processing Unit (TPU)~\cite{jouppi2018domain}. The TPU1 $\langle$TPU1$\rangle$~\cite{teich2018tearing} is only for inference, but Google soon made improvements that enabled both training and inference on the TPU2 $\langle$TPU2$\rangle$~\cite{teich2018tearing} and TPU3 $\langle$TPU3$\rangle$~\cite{teich2018tearing}.   

\item GraphCore.ai has released their C2 card $\langle$GraphCoreC2$\rangle$~\cite{lacey2017preliminary} in early 2019, which is being shipped in their GraphCore server node (see below). This company is a startup headquartered in Bristol, UK with an office in Palo Alto. They have strong venture backing from Dell, Samsung, and others. The performance values were achieved with ResNet-50 training for the single C2 card with a batch size for training of 8. The card power is an estimate based on a typical PCI card power draw. 

\item The Goya chip $\langle$Goya$\rangle$~\cite{armasu2018move,feldman2019ai} is an inference chip being developed by startup Habana Labs, which is based in San Jose and Tel Aviv. The performance was achieved on ResNet50 inference. Habana Labs is also working on a training chip called the Gaudi, which is expected to be released in mid-2019. 

\item Wave Computing has released their Dataflow Processing Unit (DPU) $\langle$WaveDPU$\rangle$~\cite{hemsoth2017first}. Each card has four DPUs. 

\item The Cambricon dataflow chip $\langle$Cambricon$\rangle$~\cite{cutress2018cambricon} was designed by a Chinese university team along with the Cambricon company, which came out of the university team. They published both int8 inference and float16 training numbers that are both significant, so both are on the chart. This is the same team that is behind the AMD Mali GPU-based Huawei Kirin chip series (see above) that are integrated into Huawei smartphones. 

\item Baidu has announced an AI accelerator chip called Kunlun $\langle$Baidu$\rangle$~\cite{merritt2018baidu, duckett2018baidu}. Presumably this chip is aimed at low power data center training and inference and is supposed to be deployed in early 2019. The two variants of the Kunlun are the 818-100 for inference and the 818-300 for training. The performance number in this chart is the Kunlun 818-300 for training. 
\end{itemize}

\subsection{Data Center Systems}
\begin{itemize}
\item There are three NVIDIA server systems on the graph: the DGX-Station, the DGX-1, and the DGX-2: The DGX-Station is a tower workstation $\langle$DGX-Station$\rangle$~\cite{alcorn2017nvidia} for use as a desktop system that includes four V100 GPUs. 
The DGX-1 $\langle$DGX-1$\rangle$~\cite{alcorn2017nvidia,cutress2018nvidias} is a server that includes eight V100 GPUs that occupies three rack units, while the DGX-2 $\langle$DGX-2$\rangle$~\cite{cutress2018nvidias} is a server that includes sixteen V100 GPUs that occupies ten rack units. The DGX-2 networks those sixteen GPUs together using a proprietary NV-Link switch. 

\item GraphCore.ai has released a Dell/EMC based server $\langle$GraphCoreNode$\rangle$~\cite{lacey2017preliminary} in early 2019, which contains eight C2 cards (see above). The performance values were achieved with ResNet-50 training on the full server with eight C2 cards. The training batch size for full server was 64. The server power is an estimate based on the components of a typical Intel based, dual-socket server with 8 PCI cards. 
\item Along with the aforementioned card, Wave Computing also released a server appliance $\langle$WaveSystem$\rangle$~\cite{hemsoth2017first,feldman2017wave}. The Wave server appliance includes four cards for a total of sixteen DPUs in the server chassis. 
\end{itemize}

\subsection{Announced Chips}

A number of other accelerator chips have been announced but have not published any performance and power numbers. These include: Intel Loihi~\cite{hemsoth2018first}, Facebook~\cite{knight2019cheaper}, Groq~\cite{morra2017groq}, Mythic~\cite{hemsoth2018mythic}, Amazon Web Services Inferentia~\cite{amazon2018inferentia}, Stanford's Braindrop~\cite{merritt2019ai}, Brainchip's Akida~\cite{wong2018brainchip,merritt2018brainchip}, Tesla~\cite{hao2019tesla}, Adapteva~\cite{olofsson2016epiphany}, Horizon Robotics~\cite{horwitz2019chinese}, Bitmain~\cite{chafkin2018chinas}, Simple Machines~\cite{tenenbaum2017computing}, Eta Compute~\cite{yoshida2018startup}, and Alibaba~\cite{yu2018alibaba}, among others. 
As performance and power numbers become available for these and other chips, they will be added in future iterations of this work. 

\section{Benchmarking}

Most of the processors in the very low power space are either research chips that were developed as proof of concepts in university research labs or they are FPGA-based solutions, also usually from university research labs. However, there are a few processors that have been commercially released. These commercial low-power accelerators are of interest for many embedded machine learning inference applications in the DoD and beyond. 
Amazon Web Services has disclosed that ''... inference actually accounts for the majority of the cost and complexity for running machine learning in production (for every dollar spent on training, nine are spent on inference).''~\cite{press2018amazon}. In this section, we will present the preliminary results of benchmarking Google TPU Edge~\cite{tpu2019edge} and Intel Movidius X-based~\cite{hruska2018nvidia} Neural Compute Stick 2 (NCS2) systems and comparing them to an Intel Core i9-9900k processor system. 

All of the benchmarks in this section were executed on an Intel-based tower desktop computer with an Intel Core i9-9900k, 32GB (3200Mhz) RAM, and a Samsung 970 Pro NVME storage disk. It was running  Windows 10 Pro (10.0.17763 Build 17763) in a VirtualBox v 6.0 virtual machine. 
The neural network model that the Google EdgeTPU ran was Mobilenet v1~\cite{howard2017mobilenets} with single shot multibox detectors (SSD)~\cite{liu2016ssd} trained with the Microsoft COCO images library~\cite{lin2014microsoft}. The model that the Intel Neural Compute Stick 2 (NCS2) and the Intel i9 9900 system ran was Mobilenet v2~\cite{sandler2018mobilenetv2} also with SSD and trained with COCO. 
The Edge TPU and NCS2 both had throttles imposed by the software that only allowed one image to be submitted for classification at a time (batch size = 1). Further, for both systems the entire neural network model had to be loaded onto the device for each image that is processed. This seems to be in place to emphasize that these are development products rather than production products, but in an actual embedded system, this limitation would not be in place since  
more performance would be gained by simultaneously submitting more than one image for classification (batch size $>$ 1), but that was not enabled or tested with this benchmarking effort. 
The NCS2 model was prepared for download to the device with the Intel Distribution of the OpenVINO (Open Visual Inference and Neural network Optimization) toolkit: 2018 R5.0.1 (30, Jan 2019). For both the TPU Edge and NCS2 devices, power draw was measured with a USB multimeter. Finally, on the Intel Core i9-9900k, TensorFlow was compiled to separately use the SSE4 and AVX2 vector engine instruction sets. The measurements for these two trials are depicted as i9-SSE4 and i9-AVX2, respectively. The Intel Core i9-9900k performs somewhat better and draws more power than typical VPX board based embedded single board computers from companies including Curtiss-Wright and Mercury Systems~\cite{curtisswright2019intel,mercury2019ciov}, which generally are based on Intel Core i7 processors that draw a maximum of 70W for the entire system. 

\begin{table}\footnotesize
  \caption{Embedded Device Descriptions}
  \label{tab:DeviceDescriptions}

\begin{tabular}{| p{0.75in} | p{0.47in}c | p{0.47in}c | p{0.47in}c | p{0.47in}c | } \hline

 & \textbf{EdgeTPU} & \textbf{NCS2} & \textbf{i9-SSE4} & \textbf{i9-AVX2} \\ \hline

NN Environment & TensorFlow Lite & OpenVINO & TensorFlow & TensorFlow \\ \hline
Mobilenet Model & v1 & v2 & v2 & v2 \\ \hline
Reported GOPS & 58.5 & 160 & & \\ \hline
Measured GOPS & 47.4 & 8.29 & 38.4 & 40.9 \\ \hline
Reported Power (W) & 2.0 & 2.0 & 205 & 205 \\ \hline
Measured Power (W) & 0.85 & 1.35 & & \\ \hline
Reported GOPS/W & 29.3 & 80.0 &  &  \\ \hline
Measured GOPS/W & 55.8 & 6.14 & & \\ \hline
Avg. Model Load Time (s) & 3.66 & 5.32 & 0.36 & 0.36 \\ \hline
Avg. Single Image Inference Time (ms) & 27.4 & 96.4 & 19.6 & 20.8 \\ \hline

\end{tabular}
\end{table}

\begin{figure}[th]
    \centering
    \includegraphics[width=3.5in]{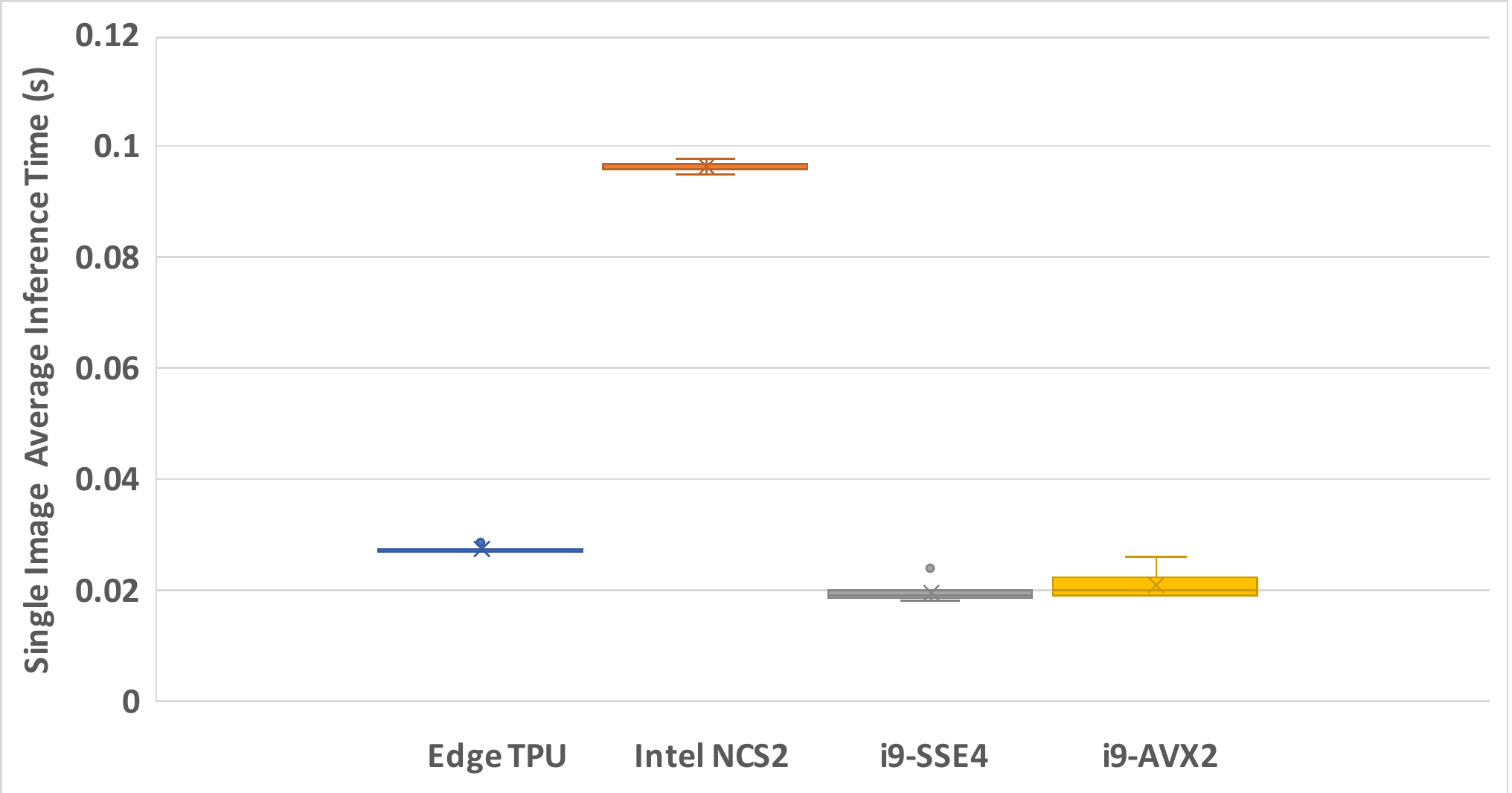}
    \caption{Box and whisker plot of single image inference times.}
    \label{fig:InferenceTimes}
  \end{figure}

Table~\ref{tab:DeviceDescriptions} summarizes the reported and measured giga operations per second (GOPS), power (W), and GOPS/W along with average model load time in seconds and average single image inference time in milliseconds. One can observe that the TPU Edge and NCS2 have much lower power consumption and much higher model load times then the Intel i9. However, single image inference times are generally the same, though the NCS2 is somewhat slower. Also, the Edge TPU GOPS/W numbers are reasonably similar, while the measured GOPS/W is much lower than the reported GOPS/W for the NCS2. Further, Figure~\ref{fig:InferenceTimes} shows a box and whiskers plot of the average and standard deviation of single image inference times for each of the four technologies. From the box and whiskers plot, we see that the single image inference times are reasonably uniform across all four technologies. 

As more low power commercial systems become available, we intend to purchase and benchmark them to add to this body of work. We expect to have performance and power numbers for the NVIDIA Jetson Xavier~\cite{hruska2018nvidia} and perhaps the NVIDIA Jetson NANO in time for the conference.

\section{Summary}

In this paper, we have presented a survey of processors and accelerators for machine learning, specifically deep neural networks along with some benchmarking results that we conducted on commercial low power processing systems that are relevant to DoD and other embedded applications. We started by overviewing the trends in machine learning processor technologies -- that many processor trends including transistor density, power density, clock frequency, and core counts are no longer increasing. This is prompting a drive to application specific accelerators that are designed specifically for deep neural networks. Several factors that determine accelerator designs were discussed including the types of neural networks, training versus inference, and numerical precision for the computations. We then surveyed and analyzed machine learning processors categorized into six regions that roughly correspond to performance and power consumption. Finally, we presented benchmarking results for two low power machine learning accelerator systems, the Google Edge TPU and the Intel Movidius X Neural Compute Stick 2 (NCS2) and compared the results to an Intel i9-9900k processor system using the SSE4 and AVX2 vector engine instruction sets.


\bibliographystyle{IEEEtran} 
\bibliography{MLProcessors}


\end{document}